\documentclass[a4paper,11pt]{article}
\usepackage{epsfig}
\usepackage{amssymb}
\usepackage{comment}
\usepackage[section]{placeins}

\usepackage[cp1251]{inputenc} 
\usepackage[T1,T2A]{fontenc}
\usepackage[english,russian]{babel}

\begin{document}


\hfill УДК 539.12, 539.171.12/6

\vspace*{\fill}

\begin{center}
{\Large\bf Calculating particle correlators with the account of detector efficiency }
\end{center}

\vspace*{\fill}

\begin{center}
{
Y.A.~Kulchitsky~$^{1),\ 2)}$,
 R.~Lednicky~$^{1),\ 3)}$,
 F.~Rimondi~$^{4)}$,
N.A.~Rusakovich~$^{1)}$,
P.V.~Tsiareshka~$^{1),\ 2)}$
 }

\end{center}

\bigskip

{\small\it
\begin{tabular}{lp{4.8in}}
 $^{1)}$ & Joint Institute for Nuclear Research,  Dubna,  Russia\\
 $^{2)}$  &  Institute of Physics, National Academy of Sciences, Minsk,  Belarus\\
 $^{3)}$  &  Institute of Physics ASCR, Prague,  Czech Republic\\
 $^{4)}$  &  Instituto  Nazionale  di  Fisica  Nucleare, Bologna,  Italy\\
\end{tabular}
}

\vspace*{\fill}

\centerline{\bf Abstract}
\smallskip
\noindent
The formulae for $m$-order correlators $K_m$ of a given particle observable (e.g.\ energy or transverse momentum) accounting for the track reconstruction efficiency are presented.  The calculation of the second- and third-order correlators is considered in some detail. Similar to the case of an ideal detector, the correlators can be expressed through the event-by-event fluctua\-tion measures of the observable single event mean and the  pseudo-central moments of the observable distribution. Howe\-ver, compared with the ideal case, this splitting allows for a lesser reduction of the computation time due to the increased number of pseudo-correlators and additional calculations of the moments of the distribution of the track weights.

\vspace*{\fill}

\newpage
\section{Introduction}

The investigation of  correlations is very important for hadron physics \cite{Kittel:2005fu,De Wolf:1995pc,Dremin:2000ep}. The integral correlation characteristics --- the correlators of particle energies, transverse momenta or rapidities --- have been suggested \cite{Manjavidze:2001ni,Sisakian:2004rv} to study the production mechanism of very  high multiplicity (VHM)  events. It was shown \cite{Amelin:2005db} that the correlators are closely related with the event-by-event fluctuations of the event mean particle observables. As a result, in the case of an ideal 100\% efficient detector, a fast and simple procedure to calculate the correlators with the help of the fluctuation measures and so called event-wise pseudo-correlators has been suggested \cite{Amelin:2005db}, exploiting the expressions of pseudo-correlators  through the central moments of the observable distribution \cite{Stadnik:2003}. The correction terms generated in the correlator analysis due to multiplicity dependent observable mean have been investigat\-ed in \cite{Filip:2007ct}. The two-particle transverse momentum correlators have been used as a correlation measure and studied as a function of event centrality in $\rm Au+Au$ collisions at RHIC \cite{Adams:2005ka}; doing this the track reconstruction efficiencies were neglected.
In this paper, we formulate the calculation procedure in the case of a finite detector efficiency.

\section{Particle correlators}

For an ideal detector, the $m$-th order correlator 
in the events with a given charged hadron multiplicity $n$ is defined as
\begin{equation} 
 \label{eq_K}
K_m (n)   =
\left\langle
\frac{1}{c^n_m}
\sum_{i_1=1}^{n- (m-1)}
\ldots
\sum_{i_m=i_{m-1}+1}^{n}
\Delta\varepsilon_{i_1}^{(l)}
\ldots\Delta\varepsilon_{i_m}^{(l)}
\right\rangle ,
\end{equation}
\begin{equation} 
 \label{eq_De}
\Delta\varepsilon_{i_\lambda}^{(l)}  = \varepsilon_{i_\lambda}^{(l)} -\langle\varepsilon \rangle .
\end{equation}

\noindent
Here  $c^n_m = \frac{n!}{m!(n-m)!}$ is the normalization factor equal to the number of combinati\-ons, $\varepsilon_{i_\lambda}^{(l)}$ 
is the observable (e.g., energy or momentum) of the $i_\lambda$-th charged hadron ($i_1 
< \ldots < i_m$) in the $l$-th event, $n$ is the charged hadron multiplicity in an event. The observable mean
\begin{equation} 
 \label{eq_K1-1}
\langle\varepsilon  \rangle =  \langle \overline{\varepsilon }^{(l)}\rangle ,
\end{equation}
where $\overline{\varepsilon}^{(l)}$ is the observable average in the $l$-th event:
\begin{equation} 
 \label{eq_K1-2}
\overline{\varepsilon}^{(l)} =
\frac{1}{n}\sum_{i=1}^{n} \varepsilon_i^{(l)}
\end{equation}
and
\begin{equation} 
 \label{eq_mean}
\left\langle  \right\rangle =\frac{1}{N(n)} \sum_{l = 1}^{N(n)}
\end{equation}
stands for the averaging over the $N (n)$ events with the charged hadron multiplicity $n$.
Note that the correlator formula (\ref{eq_K}), when formally applied to one particle, yields
$K_1 (n)=0$ according to definition of the observable means in (\ref{eq_K1-1}) and (\ref{eq_K1-2}).

Defining
\begin{equation} 
 \label{eq_De'}
\Delta\overline{\varepsilon}^{(l)}  = \overline{\varepsilon}^{(l)} -\langle\varepsilon \rangle ,
\end{equation}
\begin{equation} 
 \label{eq_De''}
\Delta\widetilde{\varepsilon}_{i}^{(l)}  = \varepsilon_{i}^{(l)} -\overline{\varepsilon}^{(l)}
\end{equation}
and using the equality
$
\Delta\varepsilon_{i}^{(l)}=\Delta\widetilde{\varepsilon}_{i}^{(l)} + \Delta\overline{\varepsilon}^{(l)} ,
$
one can decompose the correlator on the event-by-event fluctuations of the event mean observable
$ \Delta\overline{\varepsilon}^{(l)}$, event-wise pseudo-correlators $k_{\lambda}^{(l)} (n)$
and the corresponding cross terms \cite{Amelin:2005db}:
\begin{equation} 
\label{eq_Km}
K_m (n) =
\left\langle   \sum_{\lambda=0}^{m} c^m_{\lambda}  \Delta\overline{\varepsilon}^{(l)\, m-\lambda} k_{\lambda}^{(l)} (n) \right\rangle  ,
\end{equation}
where $k_0^{(l)}=1$.
The event-wise pseudo-correlators are defined similar to (\ref{eq_K}) up to the substitution
$\Delta\varepsilon_{i}^{(l)} \to \Delta\widetilde{\varepsilon}_{i}^{(l)}$:
\begin{equation} 
 \label{eq_k}
k_m^{(l)} (n)   =
\frac{1}{c^n_m}
\sum_{i_1=1}^{n- (m-1)}
\ldots
\sum_{i_m=i_{m-1}+1}^{n}
\Delta\widetilde{\varepsilon}_{i_1}^{(l)}
\ldots\Delta\widetilde{\varepsilon}_{i_m}^{(l)} .
\end{equation}
Similar to the correlator, the first order pseudo-correlator also vanishes by definiti\-on:
$k_1^{(l)}=0$. It is remarkable that the pseudo-correlators can be expressed through the central moments
of the observable distribution \cite{Amelin:2005db,Stadnik:2003}.

In the case of a non-ideal detector, one has to account for the track reconstructi\-on efficiencies
\begin{equation} 
\label{eq_w-1}
w_i^{(l)} = \frac{f_i^{(l)}}{\omega_i^{(l)} }\ ,
\end{equation}
where $\omega_i^{(l)}$ is the track reconstruction efficiency depending on pseudorapidity and $p_T$ of $i$-th track in $l$-th event,  and $f_i^{(l)}$ is a function correcting for fake, secondary and out of kinematic region tracks.
The efficiency corrected average observable in the $l$-th event is
\begin{equation} 
\label{eq_K1w-1}
\overline{\varepsilon}^{(l)} =
\frac{ \sum_{i=1}^{n}  \varepsilon_i^{(l)} w_i^{(l)} }{ \sum_{i=1}^{n}  w_i^{(l)} }.
\end{equation}
For the efficiency corrected  $m$-particle correlator, we have
{\small
\begin{equation} 
 \label{eq_Kw_m}
K_m (n)   =
\left\langle
\frac{
\sum_{i_1=1}^{n- (m-1)}
\ldots
\sum_{i_m=i_{m-1}+1}^{n}
w_{i_1}^{(l)}
\ldots w_{i_m}^{(l)}
\Delta\varepsilon_{i_1}^{(l)}
\ldots\Delta\varepsilon_{i_m}^{(l)}
}{
\sum_{i_1=1}^{n- (m-1)}
\ldots\sum_{i_m=i_{m-1}+1}^{n}
w_{i_1}^{(l)}
\ldots w_{i_m}^{(l)}
}
\right\rangle .
\end{equation}
}
The decomposition similar to (\ref{eq_Km}) now takes the form
\begin{equation} 
\label{eq_Kmr}
K_m (n) =
\left\langle   \sum_{\lambda=0}^{m} c^m_{\lambda}  \Delta\overline{\varepsilon}^{(l)\, m-\lambda} k_{\lambda}^{(l,m)} (n) \right\rangle ,
\end{equation}
where $k_0^{(l,m)}=1$. Note that now the pseudo-correlators $k_\lambda^{(l,m)}$ depend also on the correlator order $m$:
\begin{equation} 
 \label{eq_Kw_m}
k_\lambda^{(l,m)} (n)   =
\frac{
\sum_{i_1=1}^{n- (m-1)}
\ldots
\sum_{i_m=i_{m-1}+1}^{n}
w_{i_1}^{(l)}
\ldots w_{i_m}^{(l)}
\Delta\widetilde{\varepsilon}_{i_1}^{(l)}
\ldots\Delta\widetilde{\varepsilon}_{i_\lambda}^{(l)}
}{
\sum_{i_1=1}^{n- (m-1)}
\ldots
\sum_{i_m=i_{m-1}+1}^{n}
w_{i_1}^{(l)}
\ldots w_{i_m}^{(l)}
} .
\end{equation}
Obviously, such a pseudo-correlator coincides with the true one for $\lambda=m$ only:
$k_m^{(l,m)}=k_\lambda^{(l)}$.
Again, 
$K_1 = k_1^{(l)}=0$  by definition. Note, however, that the
pseudo-correlators $k_1^{(l,m)}$ don't vanish for $m > 1$.

Particularly, the second-order correlator can be decomposed as
\begin{equation} 
 \label{eq_K2b}
K_2 (n)= \left\langle
\Delta\overline{\varepsilon}^{(l)2}
+2\Delta\overline{\varepsilon}^{(l)} k_{1}^{(l,2)} (n)
+ k_{2}^{(l,2)} (n)
\right\rangle  .
\end{equation}
Here the first term
$
\left\langle
\Delta\overline{\varepsilon}^{(l)2}
\right\rangle
$
is a quadratic measure of the fluctuation of the observable event-wise mean around the sample mean.
The second term is a cross term which vanishes in the ideal case of unit reconstruction weights $w_i^{(l)}$ since the first-order event-wise pseudo-correlator
for ideal detector $k_1^{(l)}$ vanishes by definition.
Similarly, the three-particle correlator is decomposed
into four terms:
\begin{equation} 
 \label{eq_K3b}
K_3 (n)=
\left\langle
\Delta\overline{\varepsilon}^{(l)3}
+3\Delta\overline{\varepsilon}^{(l)2} k_{1}^{(l,3)} (n)
+3\Delta\overline{\varepsilon}^{(l)} k_{2}^{(l,3)} (n)
+ k_{3}^{(l,3)} (n)
\right\rangle  .
\end{equation}
Here the first term
$
\left\langle
\Delta\overline{\varepsilon}^{(l)2}
\right\rangle
$
is a cubic measure of the fluctuation of the observable event-wise mean around the sample mean.
The second and third terms are
cross terms, the first of them vanishing in the case of an ideal detector due to vanishing of the first-order event-wise pseudo-correlator
$k_1^{(l)}$.


\section{Event-wise pseudo-correlators and  central\\  moments}
We will describe in some detail the calculation of the second- and third-order pseudo-correlators.
In the case of an ideal detector, one may use the identity
\begin{equation} 
 \label{eq_s2-0}
 \sum_{i=1}^{n} \Delta\widetilde{\varepsilon}_i^{(l)}  =0
 \end{equation}
and its powers
to express the event-wise pseudo-correlators $k_m^{(l)}(n)$
through the central moments $S_\lambda^{(l)}, \lambda\le m$, of the single-particle observable distribution,
\begin{equation} 
 \label{eq_s2-1}
 S_\lambda^{(l)}(n)  = \frac{1}{n} \sum_{i=1}^{n} \Delta\widetilde{\varepsilon}_i^{(l)\lambda}.
\end{equation}
Thus, for the second- and third-order pseudo-correlator we have \cite{Amelin:2005db,Stadnik:2003}:
\begin{equation} 
 \label{eq_s2}
 k_2^{(l)}(n) = -
 \frac{1}{n-1}\cdot S_2^{(l)} ,
\end{equation}
\begin{equation} 
 \label{eq_s3}
 k_3^{(l)}(n) =
 \frac{2}{(n-1)(n-2)}\cdot S_3^{(l)} .
\end{equation}

Using the identity 
\begin{equation} 
 \label{eq_ww}
 \sum_{i=1}^{n}  \sum_{j=1, \ne i}^{n} 
w_i^{(l)}w_j^{(l)}f_{ij}=
 \sum_{i=1}^{n} w_i^{(l)} \left( \sum_{j=1}^{n} w_j^{(l)}f_{ij}-w_i^{(l)}f_{ii} \right) ,
 \end{equation}
where $f_{ij}$ is arbitrary function. After generalizing identity (\ref{eq_s2-0}) and its second power for the case of a non-ideal detector:
\begin{equation} 
 \label{eq_s2-01}
 \sum_{i=1}^{n} w_i^{(l)} \Delta\widetilde{\varepsilon}_i^{(l)}  =0 ,
 \end{equation}
\begin{equation} 
 \label{eq_s2-02}
 \sum_{i=1}^{n} (w_i^{(l)}\Delta\widetilde{\varepsilon}_i^{(l)} )^2
+ 
 \sum_{i=1}^{n}  \sum_{j=1, \ne i}^{n} 
w_i^{(l)} w_j^{(l)} \Delta\widetilde{\varepsilon}_i^{(l)}
\Delta\widetilde{\varepsilon}_j^{(l)}
=0 
 \end{equation}
and substituting the sums over the ordered $m$-plets $\{i_1<\ldots<i_m\}$
in the pseudo-correlator definitions by the sums
over the $m$-plets $\{i_1\ne\ldots\ne i_m\}$, one gets for the first- and second-order
pseudo-correlator contributing to the second-order correlator:
\begin{equation} 
 \label{eq_s2w1}
 k_1^{(l,2)}(n) = - 
\frac{\sum_{i=1}^{n} w_i^{(l)2}\Delta\widetilde{\varepsilon}_i^{(l)}}{ 
n \left( n \overline{w}^{(l) 2} - \overline{w^{(l) 2}} \right) } ,
\end{equation}
\begin{equation} 
 \label{eq_s2w}
 k_2^{(l,2)}(n)\equiv k_2^{(l)}(n) = -
\frac{\sum_{i=1}^{n} w_i^{(l) 2}\Delta\widetilde{\varepsilon}_i^{(l) 2}}{
n \left( n \overline{w}^{(l)2} - \overline{w^{(l)2}} \right) } ,
\end{equation}
where 
\begin{equation} 
 \label{eq_waver}
 \overline{w^{(l)\lambda}}=\frac{1}{n}\sum_{i=1}^n w_i^{(l)\lambda}.
 \end{equation}
Formula (\ref{eq_s2w}) shows that the pseudo-correlator $k_2^{(l)}$ is negatively defined and
doesn't explicitly depend on correlations of the observables of different particles.
Note that it can be rewritten as:
\begin{equation} 
 \label{eq_s2w-2}
 k_2^{(l)}(n) = - 
\frac{ \overline{w}^{(l)} S_2^{\prime \, (l,2)} }{ 
n\overline{w}^{(l)2} - \overline{w^{(l)2}} 
}
,
\end{equation}
where $S_2^{\prime (l)}$ is the event-wise second pseudo-central moment:
\begin{equation} 
 \label{eq_s2w-3}
 S_2^{\prime \, (l,2)}(n) =
\frac{\sum_{i=1}^{n} w_i^{(l)2} \Delta\widetilde{\varepsilon}_i^{(l)2}   }{\sum_{i=1}^{n} w_i^{(l)}}.
\end{equation}
We use the prefix  "pseudo''  because of the quadratic weights in (\ref{eq_s2w-3}) for 
$S_2^{\prime \, (l,2)}$ 
contrary to the linear weights in the true efficiency corrected 
$\lambda$-th central moment:
\begin{equation} 
 \label{eq_s2w-1}
 S_{\lambda}^{(l)}(n) = 
\frac{\sum_{i=1}^{n} w_i^{(l)} \Delta\widetilde{\varepsilon}_i^{(l) \lambda}   }{\sum_{i=1}^{n} w_i^{(l)}} .
\end{equation}
Generally, the $\lambda$-order pseudo-correlators contributing to $m$-order correlator ($\lambda \le m$) can be expressed through the $\lambda$-th  pseudo-central moments calculated with the powers $\mu\le m$ of the weights:
\begin{equation} 
 \label{eq_s2w-com}
 S_{\lambda}^{\prime \, (l; \mu)}(n) =
\frac{\sum_{i=1}^{n} w_i^{(l) \mu} \Delta\widetilde{\varepsilon}_i^{(l) \lambda}   }{\sum_{i=1}^{n} w_i^{(l)}} ,
\end{equation}
where $\mu$ is degree of corrected function.
Of course, $S_\lambda^{\prime \, (l;\mu)} = S_\lambda$ in case of an ideal detector. 

Thus, the first-order pseudo-correlator in (\ref{eq_s2w1}) can be rewritten as:
\begin{equation} 
 \label{eq_s2w-12}
k_1^{(l,2)}(n) = - 
\frac{ \overline{w}^{(l)} S_1^{\prime \, (l; 2)}}{ 
n\overline{w}^{(l)2}- \overline{w^{(l)2}} 
}
 .
\end{equation}

As for the pseudo-correlators contributing  to the the third-order correlator, using in addition the identity
\begin{eqnarray} 
 \label{eq_ww}
& 
\sum \limits_{i=1}^{n} 
\sum \limits_{j=1, \ne i }^{n} 
\sum  \limits_{k=1, \ne i, j}^{n} 
w_i^{(l)}w_j^{(l)}w_k^{(l)}f_{ijk}=   \sum \limits_{i=1}^{n} w_i^{(l)} \times
& 
\\ \nonumber
& 
\Biggl[
\sum \limits_{j=1}^{n} w_j^{(l)}
\Biggl( 
\sum \limits_{k=1}^{n} w_k^{(l)}f_{ijk}-w_i^{(l)}f_{iji}-w_j^{(l)}f_{ijj} 
\Biggr)
 -w_i^{(l)} 
\Biggl( 
\sum \limits_{k=1}^{n} w_k^{(l)}f_{iik}-2w_i^{(l)}f_{iii} 
\Biggr) 
\Biggr] 
& \nonumber
\end{eqnarray}
and the third power of identity (\ref{eq_s2-01}):
\begin{eqnarray} 
 \label{eq_s2-03}
& 
\sum \limits_{i=1}^{n} (w_i^{(l)}\Delta\widetilde{\varepsilon}_i^{(l)} )^3
+ 3
\sum \limits_{i=1}^{n} 
\sum \limits_{j=1, \ne i}^{n} 
w_i^{(l)}\Delta\widetilde{\varepsilon}_i^{(l)}
( w_j^{(l)} \Delta\widetilde{\varepsilon}_j^{(l)})^2 
& 
\\ \nonumber
& 
+ 
\sum \limits_{i=1}^{n} 
\sum \limits_{j=1, \ne i}^{n} 
\sum \limits_{k=1, \ne i,j}^{n}
w_i^{(l)} w_j^{(l)} w_k^{(l)}
\Delta\widetilde{\varepsilon}_i^{(l)}
\Delta\widetilde{\varepsilon}_j^{(l)}
\Delta\widetilde{\varepsilon}_k^{(l)}
=0 , 
& \nonumber
\end{eqnarray}
one gets the third-order pseudo-correlator:
\begin{equation} 
 \label{eq_s3w1}
 k_1^{(l,3)}(n) = 
- 2
\frac{\sum_{i=1}^{n} w_i^{(l)2} \left(n\overline{w}^{(l)}  - w_i^{(l)}\right)\Delta\widetilde{\varepsilon}_i^{(l)}}{
n \left(
n^2\overline{w}^{(l)3}- 3n \overline{w}^{(l)}\overline{w^{(l)2}} + 2 \overline{w^{(l)3}}
\right) } ,
\end{equation}
\begin{equation} 
 \label{eq_s3w2}
 k_2^{(l,3)}(n) = - 
\frac{\sum_{i=1}^{n}  w_i^{(l)2} \left(n \overline{w}^{(l)}-2w_i^{(l)}\right)\Delta\widetilde{\varepsilon}_i^{(l)2}}{
n \left(
n^2 \overline{w}^{(l)3}
-3n \overline{w}^{(l)}\overline{w^{(l)2}}
+2\overline{w^{(l)3}}
\right) } ,
\end{equation}
\begin{equation} 
 \label{eq_s3w}
 k_3^{(l,3)}(n)\equiv k_3^{(l)}(n) = 
2
\frac{\sum_{i=1}^{n} w_i^{(l)3}\Delta\widetilde{\varepsilon}_i^{(l)3}}{
n \left( 
n^2\overline{w}^{(l)3}
- 3n\overline{w}^{(l)}\overline{w^{(l)2}}
+2\overline{w^{(l)3}}
\right) } .
\end{equation}
Obviously, the pseudo-correlators in the equations  (\ref{eq_s3w1}), (\ref{eq_s3w2}) and (\ref{eq_s3w})
can be expressed through the pseudo-central moment  as
\begin{equation} 
 \label{eq_s3w1_s}
 k_1^{(l,3)}(n) =
-2
\frac{ \overline{w}^{(l)} \left( n \overline{w}^{(l)} S_1^{\prime \, (l; 2)} - S_1^{\prime \, (l; 3)} \right)}{ 
n^2 \overline{w}^{(l)3} - 3n \overline{w}^{(l)} \overline{w^{(l)2} } + 2 \overline{w^{(l)3} } 
} ,
\end{equation}
\begin{equation} 
 \label{eq_s3w2_s}
 k_2^{(l,3)}(n) = -
\frac{  \overline{w}^{(l)} \left( n \overline{w}^{(l)} S_2^{\prime \, (l; 2)} - 2 S_2^{\prime \, (l; 3)} \right)}{ 
n^2 \overline{w}^{(l)3} - 3n \overline{w}^{(l)} \overline{w^{(l)2} } + 2 \overline{w^{(l)3} }
} ,
\end{equation}
\begin{equation} 
 \label{eq_s3-01}
 k_3^{(l)}(n) = 
2
\frac{\overline{w}^{(l)} S_3^{\prime \, (l; 3)}}{ 
n^2\overline{w}^{(l)3} -3n\overline{w}^{(l)}\overline{w^{(l)2}} +2\overline{w^{(l)3}} 
}  .
\end{equation}

The generalization of the expressions for the efficiency corrected second- and third-order
pseudo-correlators to any order is straightforward.

\section{ Conclusions}

The formulae for the $m$-order correlators $K_m$ of a given particle observable (e.g.\ energy, transverse momentum or rapidity) accounting for the track reconstruction efficiency are presented with some calculation details for the case of $m=2$ and 3. Similar to the case of an ideal detector, the correlators can be expressed through the event-by-event fluctuation measures of the observable single event mean and the pseudo-central moments of the observa\-ble distribution.
The number of the terms to be calculated is however higher due to the increased number of pseudo-correlators and, in addition, the moments of the distribution of the track weights.

\bigskip
\noindent
{\Large\bf Acknowledgments}
\bigskip

\noindent
We are grateful to  J.A.Budagov and J.D.Manjavidze  for useful discussions. 
We are thankful to E.Sarkisyan-Grinbaum for usful commnets. 


\end{document}